\documentclass{article}
    \usepackage[english]{babel}
    \usepackage[margin=1in]{geometry}
    \usepackage{multicol}
    \usepackage{color}
    \usepackage{amsmath}
    \usepackage{amsfonts}
    \usepackage{background}
    \usepackage[math]{blindtext}
    \usepackage{algorithmic}
    \usepackage{graphicx}
    \usepackage{amsthm}
    \usepackage[affil-it]{authblk}

\newsavebox{\ieeealgbox}

    \SetBgScale{4}
    \SetBgColor{gray}
    \SetBgAngle{90}
    \SetBgContents{arXiv: some other text goes here}
    \SetBgPosition{-1.5,-10}
    \begin{document}

    {\centering

    {\bfseries\Large Stochastic Block Coordinate Frank-Wolfe Algorithm for Large-Scale Biological Network Alignment\bigskip}

    Yijie Wang\textsuperscript{1} and Xiaoning Qian\textsuperscript{1} \\
      {\itshape
    \textsuperscript{1} Department of Electrical \& Computer Engineering, Texas A\&M University,\\ College Station, Texas, USA, 77843 \\
    \normalfont (Dated: May 23, 2015)

      }
    }

    \begin{abstract}
With increasingly ``big'' data available in biomedical research, deriving accurate and reproducible biology knowledge from such big data imposes enormous computational challenges. In this paper, motivated by recently developed stochastic block coordinate algorithms, we propose a highly scalable randomized block coordinate Frank-Wolfe algorithm for convex optimization with general compact convex constraints, which has diverse applications in analyzing biomedical data for better understanding cellular and disease mechanisms. We focus on implementing the derived stochastic block coordinate algorithm to align protein-protein interaction networks for identifying conserved functional pathways based on the IsoRank framework. Our derived stochastic block coordinate Frank-Wolfe (SBCFW) algorithm has the convergence guarantee and naturally leads to the decreased computational cost (time and space) for each iteration. Our experiments for querying conserved functional protein complexes in yeast networks  confirm the effectiveness of this technique for analyzing large-scale biological networks.
    \bigskip


    \end{abstract}

    
    \section{Introduction}

First-order methods in convex optimization have attracted significant attention in statistical learning in recent years. They are appealing to many learning problems, such as LASSO regression and matrix completion, which have diverse applications in analyzing large-scale biological systems and high-dimensional biomedical measurement profiles~\cite{Zaslavskiy09, Klau09}. These first-order optimization methods scale well with the current ``big'' data in many biomedical applications due to their {advantages that they have low computation burden per iteration} and they are easy to be implemented on parallel computational resources.  

In this paper, we focus on the Frank-Wolfe algorithm, which is also known as the conditional  gradient method. One of its advantages is that at each iteration step it decomposes the complex constrained optimization problem into sub-problems which are easier to solve. Additionally, it is a projection free algorithm, which avoids solving the projection problem for constrained optimization as done in many other algorithms. The original Frank-Wolfe algorithm, developed for smooth convex optimization on a polytope, dates back to Frank and Wolfe~\cite{FWold}. Dunn and Harshbarger~\cite{FWg1, FWg2} have generalized the algorithm to solve the optimization for more general smooth convex objective functions over bounded convex feasible regions. Recently, researchers~\cite{BCFW} have proposed stochastic optimization ideas to scale up the original Frank-Wolfe algorithm. 

Based on these previous seminal efforts, our main contribution in this paper is that we generalize the stochastic block coordinate Frank-Wolfe algorithm proposed in~\cite{BCFW}, previously with \emph{block separable constraints}, to solve more general optimization problems with \emph{any convex compact constraints}, including the problems with block inseparable constraints. Such a generalized algorithm has a broader range of biomedical applications, including biological network alignment. We prove the convergence of our generalized stochastic block coordinate Frank-Wolfe algorithm and evaluate the algorithm performance for querying conserved functional protein complexes in real-world protein-protein interaction (PPI) networks.

In the following sections, we first describe the model formulation of the optimization problems that we are generally interested. Specifically, to address potential difficulty from more general convex compact constraints, we derive a new stochastic block coordinate Frank-Wolfe algorithm and provide the convergence proof. Then, we formulate the IsoRank problem for network alignment~\cite{isorank} as a convex programming problem and develop a SBCFW-IsoRank algorithm based on our new stochastic block coordinate Frank-Wolfe algorithm. At last, in our experiments, we show the efficiency and effectiveness of our algorithm for solving the PPI network query problem.

\section{Stochastic Block Coordinate Descent Frank-Wolfe Algorithm} \label{sec2}

Consider the minimization problem:
\begin{equation}\label{eq1}
\begin{aligned}
\textup{min:}& \ \  f(\mathbf{x})\\
s.t. & \ \  \mathbf{x} \in \mathfrak{D},
\end{aligned}
\end{equation}
where the objective function $f(\mathbf{x})$ is convex and differentiable on $\mathbf{R}^N$, and the domain $\mathfrak{D}$ is a compact convex subset of any vector space. We assume that the optimal solution $\mathbf{x}^*$ to the above problem is non-empty and bounded without loss of generality.  

Assume that we can decompose the solution space $\mathbf{R}^N$ into $n$ equal-size subspaces:
\begin{equation}
\mathbf{R}^N = \overset{n}{\underset{i=1}{\oplus}}  \mathbf{R}^{N_i}, \ \ N = \sum_{i=1}^{n} N_i,
\end{equation}
where $N_1=\ldots=N_i = \ldots, N_n$ and $\mathbf{R}^{N_i}$ denotes the $i$th equal-size subspace along the corresponding coordinates. This decomposition enables scalable stochastic optimization algorithms. Based on this decomposition, we introduce matrices $U_i$, who sum up to an identity matrix $I_N= \sum_{i=1}^n U_i$,
and $U_i$ is a $N\times N$ matrix with $U_i(t,t) = 1, t\in R^{N_i}$ on its diagonal and the other entries being equal to zero. In typical stochastic optimization algorithms~\cite{sopt}, instead of computing the gradient $\nabla f(\mathbf{x})$ at each iteration, the \emph{partial gradient} of $f(\mathbf{x})$ on a randomly selected subspace $\mathbf{R}^{N_i}$ is used: 
\begin{equation}
\nabla_i f(\mathbf{x}) = U_i\nabla f(\mathbf{x}).
\end{equation}


Now we generalize the previous stochastic block coordinate Frank-Wolfe algorithm derived in~\cite{BCFW} to solve more general optimization problems with any compact convex constraints $\mathfrak{D}$. The new generalized stochastic block coordinate Frank-Wolfe (SBCFW) algorithm is illustrated in \textbf{Algorithm 1}. In the pseudo code, the operation $i = \mathfrak{C}_k$ randomly selects one of the $n$ equal-size subspaces to update the partial gradient at each iteration with the same probability. \textcolor{black}{In addition, $U_j \times \mathbf{s} = U_j \times \mathbf{x}^k$ denotes the condition that the elements of the $j$th block of $\mathbf{s}$ equal to the elements of the $j$th block of $\mathbf{x}^k$.}

\begin{table}[ht]
\begin{center}
  \begin{tabular}{ lc }
  \hline
\textbf{Algorithm1: } Generalized SBCFW Algorithm\\
    \hline
1 \ Let $\mathbf{x}^0 \in \mathfrak{D}$, $k=0$.\\
2 \ \textbf{While} Stopping criteria not satisfied \textbf{do}\\
3 \ \ \ \textcolor{black}{Randomly divide $\mathbf{R}^N$ into $n$ blocks $\mathbf{R}^N = \overset{n}{\underset{i=1}{\oplus}}  \mathbf{R}^{N_i}$};\\
4 \ \ \ Choose $i = \mathfrak{C}_k$;\\
5 \ \ \ Find $\mathbf{s}^k_i$ such that\\
6 \ \ \ \ \ \ \ \textcolor{black}{$\mathbf{s}^k_i := arg \underset{\begin{subarray}{c} U_j \times \mathbf{s} = U_j \times \mathbf{x}^k, \forall j\ne i; \\ \mathbf{s}\in\mathfrak{D} \end{subarray}}{\textup{min}} \nabla_i f(\mathbf{x}^k)^T(\mathbf{s}-\mathbf{x}^k)$};\\
7 \  \ \ Determine the step size $\gamma$\\
8 \  \ \ \ \ \ \ \textcolor{black}{$\gamma:= arg \underset{\gamma\ \in [0,1]}{\textup{min}} f((1- \gamma)\mathbf{x}^k + \gamma \mathbf{s}^k_i)$};\\
9 \ \ \ Update $\mathbf{x}^{k+1} := (1- \gamma)\mathbf{x}^k + \gamma \mathbf{s}^k_i$;\\
10\ \ \ $k=k+1$;\\
11\ \textbf{Endwhile} \\
\hline
  \end{tabular}
  \end{center}
 \end{table}

Note that our generalized SBCFW algorithm is similar to the algorithm in~\cite{BCFW}, which aims to solve optimization problems with block separable constraints and has the sub-linear convergence property. However, our algorithm provides a more generalized framework, which can manipulate any convex and compact constraints no matter whether they are block separable or not. Because the setup of our algorithm is more general without any specific structure, it is difficult to obtain theorectical convergence rate guarantees. In this paper, we only provide the proof that our SBCFW converges to the global optimum. The convergence guarantee of the generalized SBCFW algorithm is provided by \textit{\textbf{Theorem 1}} below, which is based on \textit{\textbf{Lemma 1}}:


\noindent \textit{\textbf{Lemma 1: }} At each iteration of the SBCFW algorithm, the following inequality holds
\begin{equation}
\nabla f(\mathbf{x}^k)^T\left ( E_i[\mathbf{s}^k_i] - \mathbf{x}^k \right ) \leq 0,
\end{equation}
where $E_i[\mathbf{s}^k_i]$ is the expectation of $\mathbf{s}^k_i$ with respect to the random selection of the $i$th cordinate block to the corresponding subspace.

\begin{proof}
Assuming at the $k$th iteration, we solve the following optimization problem:
\begin{equation}\label{opt1}
\begin{aligned}
\textup{min:}& \ Z_k^i(\mathbf{s}):= \nabla_i f(\mathbf{x}^k)^T(\mathbf{s}-\mathbf{x}^k)\\
                s.t.& \ U_j \times \mathbf{s} = U_j \times \mathbf{x}^k, \ \forall j\ne i,\\
                     & \ \mathbf{s} \in \mathfrak{D}. 
\end{aligned}
\end{equation}

The solution to~\eqref{opt1} is $\mathbf{s}^k_i$. With $\mathbf{s}^k_i$ achieving the minimum of~\eqref{opt1}, we have
\begin{equation}
Z_k^i(\mathbf{s}^k_i) \leq Z_k^i(\mathbf{x}^k) = \nabla_i f(\mathbf{x}^k)^T(\mathbf{x}^k-\mathbf{x}^k) = 0.
\end{equation}
Therefore, 
\begin{equation}
 Z_k^i(\mathbf{s}^k_i)=\nabla_i f(\mathbf{x}^k)^T(\mathbf{s}^k_i-\mathbf{x}^k) \leq 0.
\end{equation}
Taking expectation on both sides of the above inequality with respect to random blocks, we obtain
\begin{equation}
\begin{aligned}
&\quad \quad \quad \quad  \quad E_i \left [ \nabla_i f(\mathbf{x}^k)^T(\mathbf{s}^k_i-\mathbf{x}^k) \right ] & \leq 0 \\
\Rightarrow & \  \quad \quad \quad \quad  \quad \dfrac{1}{n} \sum_i \nabla_i f(\mathbf{x}^k)^T(\mathbf{s}^k_i-\mathbf{x}^k) & \leq 0 \\
\Rightarrow  & \  \left ( \sum_i \nabla_i f(\mathbf{x}^k) \right )^T \dfrac{1}{n} \left ( \sum_i (\mathbf{s}^k_i-\mathbf{x}^k) \right ) &\leq 0 \\
\Rightarrow & \ \   \left ( \sum_i \nabla_i f(\mathbf{x}^k) \right )^T \left ( \dfrac{1}{n} \sum_i \mathbf{s}^k_i-\mathbf{x}^k \right ) &\leq 0 \\
\Rightarrow & \  \quad \quad \quad \quad  \nabla f(\mathbf{x}^k)^T\left ( E_i[\mathbf{s}^k_i] - \mathbf{x}^k \right ) &\leq 0.
\end{aligned}
\end{equation}
The inequality in the third line can be derived based on the fact that $\mathbf{s}^k_i - \mathbf{x}^k$ is a vector with only its $i$th coordinate block having non-zero values and the other parts being all zeros. With that, the summation in the second line can be written as the inner product between vectors $\sum_i \nabla_i f(\mathbf{x}^k)$ and $\sum_i (\mathbf{s}^k_i-\mathbf{x}^k)$.
\end{proof}

We now analyze the convergence of the new SBCFW algorithm based on \textit{\textbf{Lemma 1}}  from two cases. The first case is when
\begin{equation}\label{station}
\nabla f(\mathbf{x}^k)^T\left ( E_i[\mathbf{s}^k_i] - \mathbf{x}^k \right ) = 0.
\end{equation}
This simply means that $\mathbf{x}^k$ is a stationary point. Because the original objective function $f(\mathbf{x})$ is convex, we can conclude that $\mathbf{x}^k$ is the global minimum. Another case is when
\begin{equation}\label{station}
\nabla f(\mathbf{x}^k)^T\left ( E_i[\mathbf{s}^k_i] - \mathbf{x}^k \right ) < 0,
\end{equation}
indicating that $E_i[\mathbf{s}^k_i] - \mathbf{x}^k$ is a decent direction based on the definition~\cite{ddirect}. Hence, $E_i[\mathbf{s}^k_i] - \mathbf{x}^k$ can move along the direction to get closer to the global minimum in expectation. Furthermore, we compute the optimal step size at each iteration, therefore the objective function values are guaranteed to be non-increasing. With that, we present \textit{\textbf{Theorem 1}} as follows: 

\noindent \textit{\textbf{Theorem 1: }} The sequence $\left \{ f(\mathbf{x}^1), f(\mathbf{x}^2), ..., f(\mathbf{x}^k), ... \right \}$ generated by the SBCFW algorithm is non-increasing 
\begin{equation}
f(\mathbf{x}^1) \geq f(\mathbf{x}^2) \geq ... \geq f(\mathbf{x}^k) \geq f(\mathbf{x}^{k+1}), \ \ k \rightarrow \infty.
\end{equation}
 


\section{Biological Network Alignment}

\subsection{Optimization Model Formulation}\label{3.a}

In this section, we re-formulate the involved optimization problem for the network alignment algorithm---IsoRank~\cite{isorank} to address the potential computational challenges of aligning multiple large-scale networks. The new formulation has the same mathematical programming structure as the problem~\eqref{eq1}. 

Let $G_a$ and $G_b$ be two biological networks to align. Two networks has $N_a$ and $N_b$ vertices respectively. We define $B\in \mathbf{R}^{(N_a\times N_b)\times(N_a\times N_b)}$ as the Cartesian product network from $G_a$ and $G_b$: $B=G_a\otimes G_b$. Denote the all one vector $\mathbf{1}\in \mathbf{R}^{N_a\times N_b}$ and 
\begin{equation}
\bar{B}=B\times \textup{Diag}(B\mathbf{1})^{-1}, \label{eq:rdw}
\end{equation}
where $\textup{Diag}(B\mathbf{1})$ can be considered as a degree matrix with $B\mathbf{1}$ on its diagonal and all the other entries equal to zero. $\bar{B}$ contains the transition probabilities for the underlying Markov random walk in IsoRank~\cite{isorank}. It is well known that if $G_a$ and $G_b$ are connected networks and neither of them is bipartite graph, then the corresponding Markov chain represented by $\bar{B}$ is irreducible and ergodic, and there exists a unique stationary distribution for the underlying state transition probability matrix $\bar{B}$. The goal of the IsoRank algorithm is to find a right maximal eigenvector of the matrix $\bar{B}$: $\bar{B}\mathbf{x}=\mathbf{x}$ and $\mathbf{1}^T \mathbf{x}= 1, \ \mathbf{x} \geq 0$, which corresponds to the best correspondence relationships between vertices across two networks. When two networks are of reasonable size, spectral methods as well as power methods can be implemented to solve the IsoRank problem~\cite{isorank}. However, with large-scale networks, the transition probability matrix $\bar{B}$ can be extremely large (quadratic with $N_a\times N_b$) and spectral methods can be computationally prohibitive. In this paper, we re-formulate this problem of searching for maximal right eigenvector as a constrained optimization problem: 
\begin{equation}\label{obj}
\begin{aligned}
\textup{min:}\ & f(\mathbf{x}) := \dfrac{1}{2}\left \| \bar{B}\mathbf{x} - \mathbf{x} \right \|^2 \\
                 s.t.\ &  \mathbf{1}^T\mathbf{x} = 1, \ \mathbf{x} \geq 0. \quad\quad\quad (\mathfrak{H})
\end{aligned}
\end{equation}
After expanding the objective function, we obtain $f(\mathbf{x}) = \dfrac{1}{2}\mathbf{x}^TM\mathbf{x}$, where $M= \bar{B}^T\bar{B}-\bar{B}-\bar{B}^T+I$. Therefore the equivalent optimization problem is 
 \begin{equation}\label{obj1}
\begin{aligned}
\textup{min:}\ & f(\mathbf{x}) := \dfrac{1}{2}\mathbf{x}^TM\mathbf{x} \\
                 s.t.\ &  \mathbf{1}^T\mathbf{x} = 1, \ \mathbf{x} \geq 0. \quad\quad\quad (\mathfrak{H})
\end{aligned}
\end{equation}
The gradient of $f(\mathbf{x})$ can be easily computed $\nabla f(\mathbf{x}) = M\mathbf{x}$. Furthermore, we find that the Hessian matrix of $f(\mathbf{x})$ is $M$, which is a positive semi-definite matrix proven by \textit{\textbf{Lemma 2}}: 


\noindent \textit{\textbf{Lemma 2}}: $M= \bar{B}^T\bar{B}-\bar{B}-\bar{B}^T+I$ is positive semi-definite.

\begin{proof} 
$M$ can be written as $M=(\bar{B}-I)^T(\bar{B}-I)$, which proves the lemma.
\end{proof}

With \textit{\textbf{Lemma 2}}, it is obvious that the objective function $f(\mathbf{x})$ is convex. Also, the constraint set $\mathfrak{H} = \{\mathbf{x} | \mathbf{x}^T\textbf{1}=1, \mathbf{x}\geq 0\}$ is a unit simplex, which is convex and compact. Hence, the IsoRank problem~\eqref{obj} has the same problem structure as~\eqref{eq1} and our generalized SBCFW algorithm can be used to solve~\eqref{obj1} with much better scalability and efficiency due to the efficiency of the randomized partial gradient computation at each iteration. Similarly as in~\cite{isorank}, in addition to network topology, we can incorporate other information in the formulation for more biologically significant alignment results by replacing $\bar{B}$ with $\hat{B} = \alpha\bar{B}+(1-\alpha)\bar{S}\mathbf{1}^T, \ \alpha\in[0,1]$. Here $\bar{S} = S/ | S|$ is a normalized similarity vector with size $N_a\times N_b$, cancatenated from the doubly indexed similarity estimates $S([u,v])$ based on the sequence or function similarity between vertices $u$ in $G_a$ and $v$ in $G_b$. 


\subsection{SBCWF-IsoRank Algorithm}
As shown in Section~\ref{3.a}, $f(\mathbf{x})$ in~\eqref{obj} is convex and the constraint set $\mathfrak{H}$ in~\eqref{obj1} is a convex compact set. Therefore, we can apply the generalized SBCWF algorithm proposed in Section~\ref{sec2} to solve the corresponding optimization problem~\eqref{obj1}.  The detailed algorithm is illustrated in~\textbf{Algorithm~2}. Here we want to emphasize that, in each iteration of our SBCFW-IsoRank algorithm, both the time complexity and the space complexity are $O\left (\dfrac{N^2}{n} \right )$, which is achieved through tracking the vectors of $\mathbf{p}_k=E\mathbf{x}^k$ and $\mathbf{q}_k=E\mathbf{s}^k_i$ at step 2 and 10 of each iteration in~\textbf{Algorithm~2}, respectively. The stopping criterion is $\left \| \bar{B}\mathbf{x} - \mathbf{x} \right \| \leq \xi \left \|\mathbf{x} \right \|$, which can be efficiently estimated by
\begin{equation}
\left \| \bar{B}\mathbf{x} - \mathbf{x} \right \| = \mathbf{x}^TM\mathbf{x} = (E\mathbf{x})^TE\mathbf{x} = \mathbf{p}_k^T\mathbf{p}_k,
\end{equation}
which is taken in line 11 in the SBCFW-IsoRank algorithm.

\begin{table}[ht]
\begin{center}
  \begin{tabular}{ lc }
  \hline
 \textbf{Algorithm 2: } SBCFW-IsoRank Algorithm \\
 \hline
\textbf{Input:}  $\xi$, $n$ and $E$\\
1  \textbf{for} $k = 0, ..., \infty$ \textbf{do}\\
2  \ \ randomly divide $\mathbf{R}^N$ into $n$ equal-size parts\\
 3  \ \ choose $i = \mathfrak{C}_k$\\
 4  \ \ \textbf{if}($k==0$)\\
 5  \ \ \ \ initialize the $i$th block of $\mathbf{x}^0$ with $\dfrac{n}{N}$\\
 6  \ \ \textbf{endif}\\
 7 \ \ compute $\mathbf{p}_k=E\mathbf{x}^k$ and $\nabla_i f(\mathbf{x}^k) = [E^T]_i\mathbf{p}_k$\\
 8 \ \  solve the sub-problem: \\
 9 \ \ \ \ \ \ \ $\mathbf{s}^k_i := arg \underset{\begin{subarray}{c} U_j\times\mathbf{s} = U_j\times\mathbf{x}^k, \ \forall j\ne i;\\ \mathbf{s}\in\mathfrak{H}, \end{subarray}}{\textup{min}} \nabla_i f(\mathbf{x}^k)^T(\mathbf{s}-\mathbf{x}^k)$\\
 10  \ compute $\mathbf{q}_k=E\mathbf{s}^k_i$\\
 11 \ \textbf{if} $\mathbf{p}_k^T\mathbf{p}_k < \xi \left \| \mathbf{x} \right \|$\\
 12 \ \ \ \ \textbf{break};\\
 13 \ \textbf{endif}\\
 14 \ compute the step size $\gamma_k^*$:\\
 15 
      \ \ \ \ \ \ \ $\gamma_k^*=\left\{\begin{matrix}
\textup{min}\left \{\hat{\gamma}, 1\right \} & \hat{\gamma} > 0, \hat{\gamma} = \dfrac{\mathbf{p}_k^T\mathbf{p}_k-\mathbf{p}_k^T\mathbf{q}_k}{\mathbf{p}_k^T\mathbf{p}_k-2\mathbf{p}_k^T\mathbf{q}_k+\mathbf{q}_k^T\mathbf{q}_k}\\ 
0 & o.w.
\end{matrix} \right.$\\
 16 \ \ $\mathbf{x}^{k+1} = \mathbf{x}^k + \gamma^k(\mathbf{s}_i^k - \mathbf{x}^k)$\\
 17 \textbf{endfor}\\
 \textbf{Output:} $\mathbf{x}^k$\\
\hline
  \end{tabular}
  \end{center}
 \end{table}

\subsection{Initialization}
In order to guarantee both the time and space complexity to be $O (\frac{N^2}{n}  )$ at each iteration, we can not initialize the algorithm with randomly generated $\mathbf{x}^0$ to avoid a multiplication of a matrix of size $N\times N$ and a vector of size $N$, whose time and space complexity would be $O(N^2)$. We propose to initialize $\mathbf{x}^0$ in the following way:~First, randomly divide $\mathbf{R}^N$ into $n$ parts with equal sizes and randomly pick the $i$th part. Then, we initialize every elements in the $i$th part with $\frac{n}{N}$, which makes $\mathbf{x}^0$ in the feasible space defined by the constraint set $\mathfrak{H}$. Using the above initialization strategy, the time and space complexity for computating $\nabla_i f(\mathbf{x}^0)$,  $\mathbf{p}_0=E\mathbf{x}^0$ and $\mathbf{q}_0=E\mathbf{s}^0$ are all under $O (\frac{N^2}{n})$, which is easy to verify. 



\subsection{Algorithm to Solve the Sub-problem}

As shown in the SBCFW-IsoRank algorithm, at each iteration we need to solve a sub-problem. Fortunately, the sub-problem can be solved in a straightforward manner for the optimization problem~\eqref{obj1}. For the following sub-problem at iteration $k$: 
\begin{equation}\label{submin}
\begin{aligned}
\textup{min:}\ & \nabla_i f(\mathbf{x}^k)^T(\mathbf{s}-\mathbf{x}^k)\\
                 s.t.\ &  \mathbf{s}\in\mathfrak{H}, \\ 
                 &U_j\times \mathbf{s}= U_j\times \mathbf{x}^k, \ \forall j\ne i,
\end{aligned}
\end{equation}
the optimal solution is $\mathbf{s}^* = \mathbf{x}^k - U_i\mathbf{x}^k + L\mathbf{e}_j$, where $\mathbf{e}_j$ is an all-zero vector except that the $j$th element is 1 and $L = \sum_{l\in R^{N_i}} \mathbf{x}^k(l)$. Here, $j$ is the index of the coordinate with the smallest value in the $i$th block of $\nabla_i f(\mathbf{x}^k)$: 
 \begin{equation}
j = arg\underset{l\in \mathbf{R}^{N_i}}{\textup{min:}}\ [\nabla_i f(\mathbf{x}^k)](l).
\end{equation}

\subsection{Optimal Step Size}
To obtain the optimal step size at each iteration, we need to solve the following optimization problem: 
\begin{equation}\label{qf}
\begin{aligned}
\textup{min:}\ & \left (\mathbf{x}^k + \gamma (\mathbf{s}^k - \mathbf{x}^k ) \right )^TM\left (\mathbf{x}^k + \gamma (\mathbf{s}^k - \mathbf{x}^k ) \right )\\
s.t. \ & 0\leq \gamma \leq 1,
\end{aligned}
\end{equation}
which is the classic quadratic form with respect to $\gamma$.  If $\hat{\gamma}=\dfrac{\mathbf{p}_k^T\mathbf{p}_k-\mathbf{p}_k^T\mathbf{q}_k}{\mathbf{p}_k^T\mathbf{p}_k-2\mathbf{p}_k^T\mathbf{q}_k+\mathbf{q}_k^T\mathbf{q}_k}>0$, which is the solution to~\eqref{qf} without any constraints, the optimal solution $\gamma^*$ is the minimum value between 1 and $\hat{\gamma}$, otherwise $\gamma^*=0$.  The definition of $\mathbf{p}_k$ and $\mathbf{q}_k$ are given in lines 7 and 10 in~\textbf{Algorithm~2}.

\subsection{Time and Space Complexity}\label{IIIC}
At each iteration, the most computationally expensive operations are the updates of $\mathbf{p}_k$ and $\mathbf{q}_k$ (lines 7 and 10 of SBCFW-IsoRank) and the calculation of the partial gradient $\nabla_i f(\mathbf{x}^k)$ (line 7 of SBCFW-IsoRank). 

The calculation of $\mathbf{p}_k$ and $\mathbf{q}_k$ are similar. From line 10 of \textbf{Algorithm 2}, we know 
\begin{equation}
\begin{aligned}
\mathbf{p}_k&=E\mathbf{x}^k\\
&=E\left ( \mathbf{x}^{k-1} + \gamma^{k-1}(\mathbf{s}^{k-1}-\mathbf{x}^{k-1}) \right )\\
&=\mathbf{p}_{k-1} + \gamma^{k-1}E(\mathbf{s}^{k-1}-\mathbf{x}^{k-1}).
\end{aligned}
\end{equation}
The second equation is derived by replacing $\mathbf{x}^k$ with the equation in line 16 of our SBCFW-IsoRank algorithm. Because we keep tracking $\mathbf{p}_k$ at each iteration, we do not need to recompute $\mathbf{p}_{k-1}$. Therefore, we only need to compute $E(\mathbf{s}^{k-1}-\mathbf{x}^{k-1})$, which takes $O\left (\dfrac{N^2}{n} \right )$ operations because $(\mathbf{s}^{k-1}-\mathbf{x}^{k-1})$ is a vector, with only its $i$th block being non-zeros and all the other parts are zeros. Additionally, the memory consumption is also $O\left (\dfrac{N^2}{n} \right )$ by the similar argument. Similarly, we can compute $\mathbf{q}_k$
\begin{equation}
\begin{aligned}
\mathbf{q}_k&=E\mathbf{s}^k\\
&=E\left ( \mathbf{x}^{k} + (L\mathbf{e}_j-U_i\mathbf{x}^{k}) \right )\\
&=\mathbf{p}_{k} + E(L\mathbf{e}_j-U_i\mathbf{x}^{k}),
\end{aligned}
\end{equation}
where $(L\mathbf{e}_j-U_i\mathbf{x}^{k})$ is also a vector with only the $i$th block having non-zero values. Therefore, the computation of $\mathbf{q}_k$ also takes $O\left (\dfrac{N^2}{n} \right )$ operations and consumes $O\left (\dfrac{N^2}{n} \right )$ memory. 

The equation of calculating $\nabla_i f(\mathbf{x}^k)$ is as follows: 
\begin{equation}
\nabla_i f(\mathbf{x}^k) = [E^T]_i \mathbf{p}_k,
\end{equation}
where the operator $[\cdot]_i$ is to get the rows of the matrix corresponding to the $i$th coordinate block. Hence, it is easy to verify that the time complexity and space complexity of computing $\nabla_i f(\mathbf{x}^k)$ are $O\left (\dfrac{N^2}{n} \right )$.

In summary, based on the above analyses, both the time complexity and space complexity of our SBCFW-IsoRank at each iteration are $O\left (\dfrac{N^2}{n} \right )$.

\section{Experiments}

In this section, we apply our SBCFW-IsoRank algorithm to two network query problems. For the first set of experiments, we take a known protein complex in an archived yeast protein-protein interaciton (PPI) network in one database~\cite{Krogan06} as the query to search for the subnetwork in another yeast PPI network~\cite{hasty} with different archived interactions. We call this yeast-yeast network query problem. The goal of this set of experiments is to check the correctness of our algorithm as we have the ground truth for the target subnetwork. With that, we aim to test the convergence property of our algorithm under different partitions and the relationship between the number of iterations and number of partitions. The second experiment is to query a large-scale yeast PPI network in IntAct~\cite{intact} to find similar subnetworks of proteins with similar cellular functionalities for a known protein complex in human PPI network. The aim of this experiment is to show that our new algorithm can help transfer biology knowledge from model organisms to study potential functionalities of molecules in different organisms. 



\subsection{Yeast-Yeast Network Query Problem}

\begin{figure}[tb]
\centering
\centerline{\includegraphics[width=12cm]{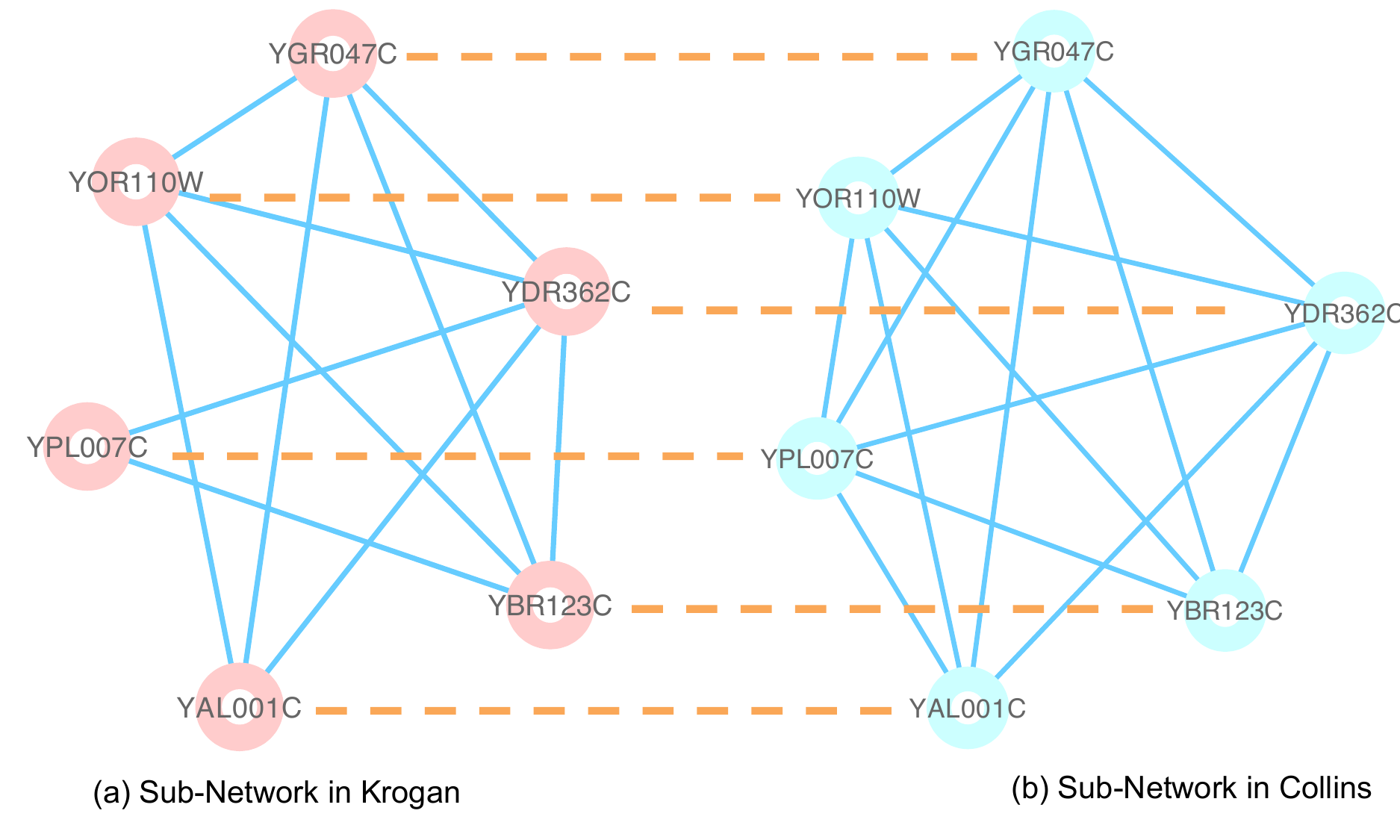}} 
\caption{Query subnetwork and its aligned result in the target network: (a) the query subnetwork in the Krogan's yeast PPI network; (b) the aligned result in Collins yeast PPI network.}
\label{fig1}
\end{figure}

We test our SBCFW-IsoRank algorithm on the yeast-yeast PPI network query problem by solving the optimization problem introduced in the previous section. We take a subnetwork with 6 proteins (Fig.~\ref{fig1}(a)) from the Krogan's yeast PPI network~\cite{Krogan06} as the query example to search for the conserved functional complex in a target network, which is the Collins network~\cite{hasty} with 1,622 proteins and 9,074 interactions. The query subnetwork is the transcription factor TFIIIC complex in the Krogan's network and we are interested in testing whether we can find the same subnetwork in the Collins network. The dimension of our optimization problem is $6\times 1,622=9,732$. We run this preliminary example so that we can compare our stochastic optimization results with the results from the power method, which is typically done in the original IsoRank algorithm~\cite{isorank}. 
Theoretically, the time and space complexity of our SBCFW-IsoRank at each iteration are both  $O(N^2/n)$ based on the analysis in Section~\ref{IIIC}. Compared to $O(N^2)$ time and space complexity for the power method by IsoRank~\cite{isorank}, our SBCFW-IsoRank algorithm can scale better with properly selected $n$. 

As both the query example and the target network contain interactions among proteins from the same organism---yeast, we can easily check the correctness of the query result. We define the accuracy as the number of corrected aligned proteins divided by the total number of proteins in the query subnetwork. We implement the  SBCFW-IsoRank algorithm for different numbers of partitions $n$ but use the same stopping criterion: $\left \| \hat{B}\mathbf{x} -\mathbf{x}\right \| \leq \xi \left \| \mathbf{x}\right \|, \xi=0.1$. In Table~I, we find that our stochastic optimization algorithm obtains the same biologically meaningful results as the power method.  

\begin{figure}[tb]
\centering
\centerline{\includegraphics[height=8cm, width=12cm]{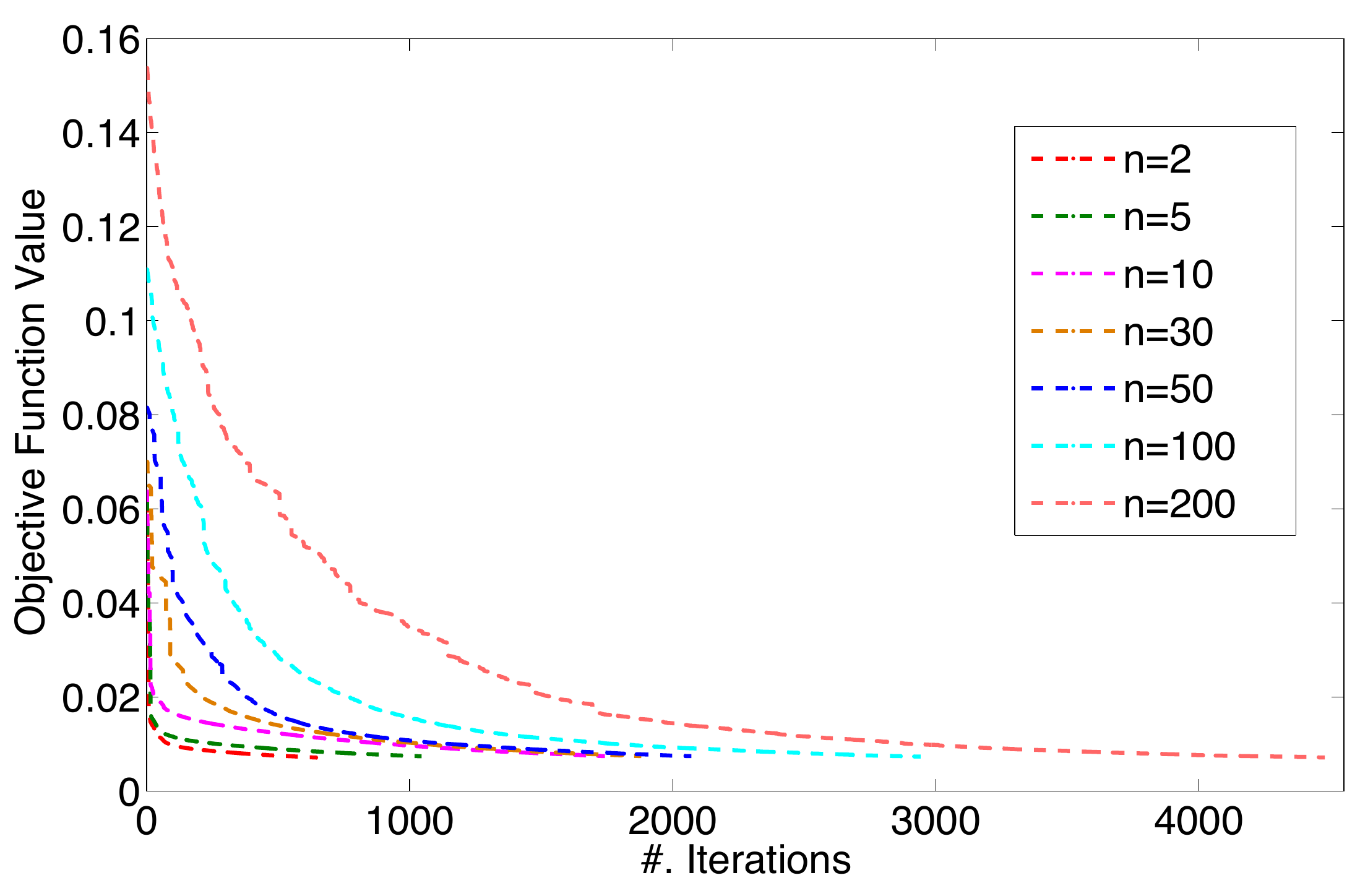}} 
\caption{The change of the objective function values with the increasing number of iterations with different numbers of partitions.}
\label{fig2}
\end{figure}

Fig.~\ref{fig2} shows the changes of the objective function values with respect to the increasing number of iterations. As illustrated in~Fig.~\ref{fig2}, our algorithm converges for all different $n$. Additionally, we find the larger the number of partitions $n$ is, the larger the number of iterations we need to have the algorithm converge to the global optimum with the same stopping criterion. This clearly demonstrates the tradeoff between the efficiency and scalability of the stochastic optimization algorithms. Interestingly, we notice that for $n=10, 30,$ and $50$, the number of iterations does not increase much, which indicates that we may achieve fast computation with reasonably large $n$ because our algorithm is more efficient for larger $n$ at each iteration.   

\begin{table}[h]
\begin{center}
\caption{Comparison on different decompositions with $\xi=0.1$.} \label{Table1Label}
\begin{tabular}{|c|c|c|c|}
 \hline
 \#. of partitions ($n$) & Computational time (s) & \#. Iterations & Accuracy\\
 \hline
 2 & 11.60 & 648 & 100\%\\
 \hline
 5 & 8.53 & 1,045 & 100\%\\
 \hline
10 & 7.44 & 1,742 & 100\%\\
 \hline
 30 & 5.05 & 1,880 & 100\%\\
 \hline
 50 & 4.93 & 2,070 & 100\%\\
 \hline
100 & 7.06 & 2,942 &100\%\\
 \hline
 200 & 13.05 & 4,478 & 100\%\\
 \hline
\end{tabular}
\end{center}
\end{table}

To further investigate the performance with different $n$, we run our algorithm 10 times for each $n$ and show the average computational time, the average number of iterations, and the average accuracy score in Table~I. From Table~I, we observe that for all different $n$, our algorithm can obtain $100\%$ accuracy, which again demonstrates the effectiveness and convergence of our generalized SBCFW algorithm. Also, we notice that with the increasing $n$, the number of iterations increase; however, the computational time is first reducing then increasing. For example, when $n=2$, our algorithm converges with the smallest number of iterations, but the computational time is not the best because at each iteration the algorithm takes $O\left ( \dfrac{N^2}{2} \right )$ operations. In contrast, when $n=50$, though the number of iterations is larger, but it reaches the global optimum with the least computation time, which is indeed twice faster than $n=2$. The trend of the computational time implies that there may exist the best number of partitions $n^*$. Empirically, when $n<n^*$ the computational time decreases while the computational time can increase when $n>n^*$. However, it is difficult to provide a theoretical proof for this observed phenomenon. Finally, for the scalibility of the algorithm, we always prefer larger $n$ to make the memory requirement as low as possible. 






\subsection{Human-Yeast Network Query Problem}

\begin{figure}[tb]
\centering
\centerline{\includegraphics[width=12cm]{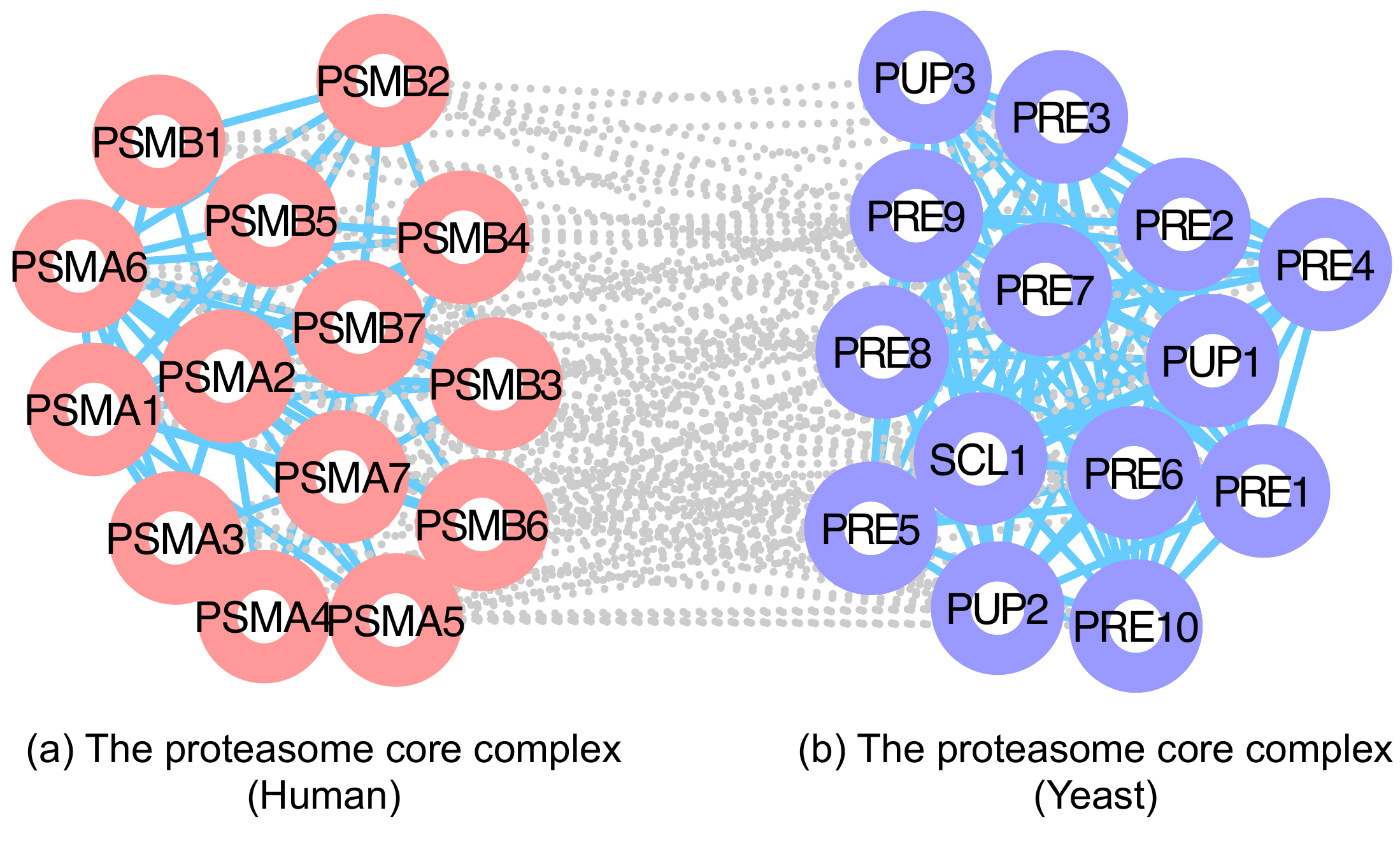}} 
\caption{Querying human protein complex in a yeast PPI network. The proteins are annotated by their gene names. The solid lines are protein interactions and the dash lines denote orthologous relationships based on protein sequence similarity by BLAST between the proteins in different organisms. (a) Human proteasome core complex; (b) The aligned proteasome core complex in yeast found by SBCFW-IsoRank.}
\label{fig3}
\end{figure}

We further study the biological signficance of network query results by our SBCFW-IsoRank algorithm. We extract a subnetwork as a query example from a human PPI network archived in~IntAct~\cite{intact}. The query subnetwork is the proteasome core complex, with induced interactions among the corresponding proteins from~IntAct~\cite{intact}. The proteasome core complex in human consists of 14 proteins in total, as shown in~Fig.~\ref{fig3}(a). The target network is the yeast PPI network, also obtained from IntAct~\cite{intact}, which has 6,392 proteins and 77,065 interactions. Our goal is to find the most similar subnetwork to the human proteasome core complex in the target yeast PPI network, based on both the interaction topology and the protein sequence similarity, which is computed by BLAST~\cite{blast}.

We first construct the alignment network, which has $N=14\times 6,392=89,488$ vertices. By our SBCFW-IsoRank algorithm with $n=300$, instead of operating a matrix of size $89,488\times 89,488$ by the power method, we only need to handle a matrix of size $298\times 89,488$. At each iteration, the computational time as well as the memory requirement are reduced 300 times. Our Matlab implementation of SBCFW-IsoRank on a MacPro notebook with 8GB RAM takes only around 750 seconds to converge by reaching the stopping criteria ($\xi=0.1$).  


The identified subnetwork in the target yeast PPI network by our algorithm is illustrated in~Fig.~\ref{fig3}(b). To evaluate the biological significance of the obtained subnetwork, we check the p-value based on GO (Gene Ontology) enrichment analysis using GOTerm~Finder~\cite{bauer}. The identified subnetwork is significantly enriched in GO term GO:0005839, which is in fact the same proteasome core complex, with p-value $9.552e-36$. This experiment demonstrates that our algorithm can find the biologically consistent groups of proteins with the same cellular functionalities as the proteins in the query subnetwork, hence with the capability of transferring existing biology knowledge in model organisms (yeast for example) to less studied organisms when the group of proteins in the query subnetwork require better understanding of their cellular functionalities.  

\section{Conclusions}

In this paper, we generalize the block coordinate Frank-Wolfe algorithm to solve general convex optimization problems with any convex and compact constraint set. Our generalized SBCFW algorithm has the convergence guarantee. We re-formulate the IsoRank problem to such a convex programming problem and solve the biological network alignment problem by our SBCFW-IsoRank algorithm, which scales better with the size of networks under study. The scalability, efficiency, and effectiveness of our algorithm on solving IsoRank are demonstrated for real-world PPI network query problems. In our future work, we will consider the derivation of the optimal partition number for better tradeoff between computational efficiency and scalability.


\section{Acknowledgements}
The authors would like to thank Simon Lacoste-Julien for pointing out the error in the original conference paper. This work was partially supported by Awards \#1447235 and \#1244068 from the National Science Foundation; as well as Award R21DK092845 from the National Institute Of Diabetes And Digestive And Kidney Diseases, National Institutes of Health.    
    
    \bibliographystyle{acm}
\bibliography{xqian}

    \end{document}